\documentclass[conference]{IEEEtran}
\IEEEoverridecommandlockouts

\usepackage{cite}
\usepackage{amsmath,amssymb,amsfonts}
\usepackage{algorithmic}
\usepackage{graphicx}
\usepackage{textcomp}
\usepackage{xcolor}
\usepackage{booktabs}
\usepackage{bbm}
\usepackage{tikz}
\usetikzlibrary{positioning,arrows.meta}

\usepackage{tabularx}
\usepackage{booktabs}
\usepackage{makecell}

\usepackage{tikz}
\usepackage{graphicx}
\usepackage{xcolor}
\usetikzlibrary{arrows.meta,positioning,calc,backgrounds}

\definecolor{frozenFill}{HTML}{EAF2FB}
\definecolor{qformerFill}{HTML}{DFF3E3}
\definecolor{loraFill}{HTML}{EFE4FF}
\definecolor{lossFill}{HTML}{FFF3BF}
\definecolor{stageFill}{HTML}{F8FAFC}
\definecolor{stageStroke}{HTML}{496A7A}
\definecolor{sColor}{HTML}{2B6CB0}
\definecolor{oColor}{HTML}{2F855A}
\definecolor{apColor}{HTML}{C05621}
\definecolor{darkLine}{HTML}{263238}

\tikzset{
  >=Latex,
  every node/.style={font=\scriptsize},
  box/.style={
    draw=darkLine,
    rounded corners=2pt,
    align=center,
    inner sep=2.2pt,
    minimum height=5.8mm,
    line width=.45pt
  },
  frozen/.style={box, fill=frozenFill, minimum width=15mm},
  qformer/.style={box, fill=qformerFill, minimum width=20mm},
  lora/.style={box, fill=loraFill, minimum width=16mm},
  loss/.style={box, fill=lossFill, minimum width=20mm, inner sep=1.8pt},
  stage/.style={
    draw=stageStroke,
    rounded corners=5pt,
    fill=stageFill,
    line width=.65pt
  },
  qS/.style={
    box,
    fill=sColor!10,
    draw=sColor,
    text=sColor,
    minimum width=10mm,
    minimum height=4.8mm,
    inner sep=1.1pt
  },
  qO/.style={
    box,
    fill=oColor!10,
    draw=oColor,
    text=oColor,
    minimum width=10mm,
    minimum height=4.8mm,
    inner sep=1.1pt
  },
  qAP/.style={
    box,
    fill=apColor!12,
    draw=apColor,
    text=apColor,
    minimum width=10mm,
    minimum height=4.8mm,
    inner sep=1.1pt
  },
  core/.style={
    box,
    fill=white,
    minimum width=18mm,
    minimum height=15mm,
    inner sep=1.4pt
  },
  flow/.style={-Latex, draw=darkLine, line width=.55pt},
  kvflow/.style={-Latex, draw=darkLine, line width=.50pt, dashed},
  sFlow/.style={-Latex, draw=sColor, line width=.80pt},
  oFlow/.style={-Latex, draw=oColor, line width=.80pt},
  apFlow/.style={-Latex, draw=apColor, line width=.80pt},
  note/.style={font=\tiny, align=center, text=darkLine!80}
}

\def\BibTeX{{\rm B\kern-.05em{\sc i\kern-.025em b}\kern-.08em
    T\kern-.1667em\lower.7ex\hbox{E}\kern-.125emX}}
\begin{document}

\title{JYVA at BeTraC: Audio-Based Medical Note Generation with SOAP Section-Specific Queries
}

\author{
\IEEEauthorblockN{Youness Dkhissi\textsuperscript{*}}
\IEEEauthorblockA{\textit{Orange Research} \\
Cesson-Sévigné, France\\
youness.dkhissi@orange.com}
\and
\IEEEauthorblockN{Valentin Vielzeuf\textsuperscript{*}}
\IEEEauthorblockA{\textit{Orange Research} \\
Cesson-Sévigné, France\\
valentin.vielzeuf@orange.com}
\and
\IEEEauthorblockN{Antoine Caubrière}
\IEEEauthorblockA{\textit{Orange Research} \\
Cesson-Sévigné, France\\
antoine.caubriere@orange.com}
\and
\IEEEauthorblockN{Jean-Emile Dartois}
\IEEEauthorblockA{\textit{Orange Research} \\
Cesson-Sévigné, France\\
jeanemile.dartois@orange.com}
\thanks{\textsuperscript{*}Equal contribution}
}
\maketitle

\begin{abstract}
We present our submission to the Beyond Transcription Challenge (BeTraC) on medical note generation from doctor-patient audio conversations. 
We propose a section-aware approach based on Qwen2.5-Omni-3B, in which a lightweight Q-Former learns SOAP section-specific queries to extract compact clinical representations from speech.
In the lightweight track, our system ranks fourth and substantially outperforms the challenge baseline, achieving a Concept F1 of 0.5005 on the official test set.
The organizers' analysis and our heuristic evaluation suggest improved robustness to hallucinations, and better stability under difficult audio conditions.
These results show that explicitly modeling SOAP structure through section-specific queries is an effective way to improve the faithfulness and reliability of audio-based medical note generation.
\end{abstract}

\begin{IEEEkeywords}
speechLLM, SOAP, speech summarization, dialogue, LLM
\end{IEEEkeywords}

\vspace{-0.1cm}
\section{Introduction} 

Large language models (LLMs) have recently achieved strong results in summarization, but factual faithfulness remains a major challenge~\cite{zhang2023benchmarkinglargelanguagemodels}.
In particular, LLMs are still prone to hallucinations, which we use here to denote generated content that is unsupported by the input, inconsistent with it, or factually incorrect~\cite{vandeemter2024pitfallsdefininghallucination,sahoo2024comprehensivesurveyhallucinationlarge}. 
Such errors are especially problematic in summarization, where outputs are expected to be not only concise and coherent, but also faithful to the source~\cite{goyal2021annotatingmodelingfinegrainedfactuality}.
These issues are particularly pronounced in the context of spoken conversations summarization.
Compared with written text, speech is typically less structured and contains disfluencies such as hesitations, repetitions, and false starts, which complicate content selection and faithful generation~\cite{shang2024endtoendspeechsummarizationusing,retkowski2025summarizingspeechcomprehensivesurvey}. 
Recent studies further show that hallucinations also affect speech-oriented generative models, including speech foundation models and SpeechLLMs~\cite{frieske2024hallucinationsneuralautomaticspeech,atwany2025losttranscriptiondistributionshift,waldendorf2026detectinghallucinationsspeechllmsinference}.

This issue is especially critical for medical conversation summarization and SOAP note generation (Subjective, Objective, Assessment and Plan), where reliability and factual accuracy are essential~\cite{carandang-etal-2025-llms}. 
Moreover, recent work suggests that evaluating hallucinations in medical note generation is itself non-trivial, as clinically appropriate abstractions or reformulations may be incorrectly penalized by surface-level matching criteria~\cite{vachhani2026literalsummarizationredefininghallucination}.
These challenges suggest that SOAP note generation should not be approached as generic free-form summarization. 
Because SOAP notes follow a predefined section structure, with each section requiring different types of evidence from the source conversation, explicitly modeling this structure may help improve both information selection and factual faithfulness.

This perspective is well aligned with the Beyond Transcription Challenge (BeTraC), which promotes end-to-end speech-to-summary systems for clinical note generation.
By encouraging models to generate SOAP notes directly from speech, rather than relying on cascaded ASR-based pipelines, BeTraC provides a relevant setting for studying how to improve the faithfulness and reliability of clinical note generation from spoken interactions.

In this work, we address the task of generating SOAP notes from medical conversations through two main contributions. 
First, we propose a section-aware approach inspired by the Q-Former paradigm~\cite{dai2023instructblip}, in which learnable queries are specialized for each SOAP section to be generated. 
This design introduces a simple yet effective inductive bias toward section-specific information extraction and generation. 
Second, through both the official BeTraC evaluation and our complementary heuristic analyses, we show that this section-aware design improves not only task performance, but also the factual reliability of the generated notes.

\vspace{-0.1cm}
\section{Methodology} 

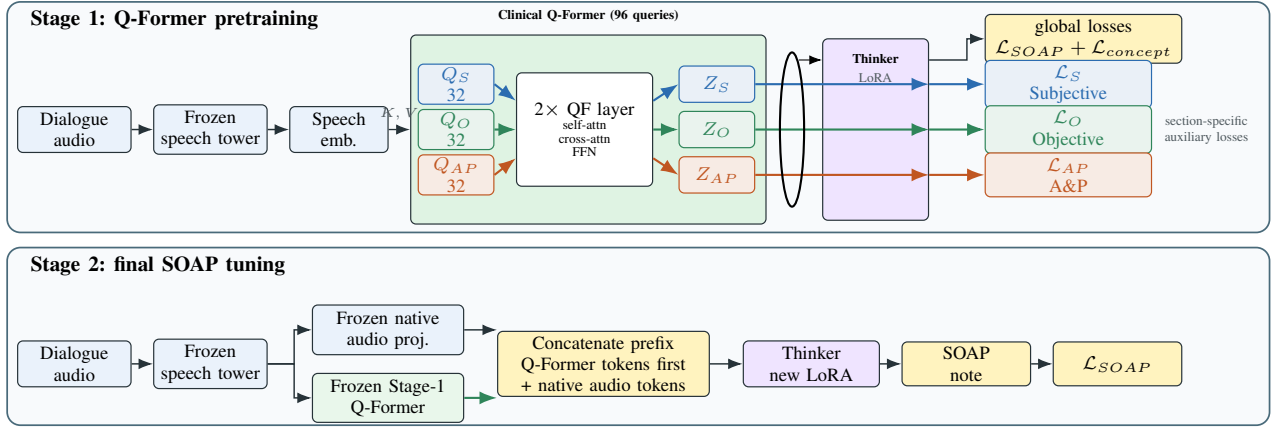
\begin{figure*}[t]
\centering
\begin{tikzpicture}[x=1cm,y=1cm]

\begin{scope}[on background layer]
  \node[
    stage,
    anchor=north west,
    minimum width=16.70cm,
    minimum height=3.05cm
  ] at (-0.15,6.05) {};

  \node[
    stage,
    anchor=north west,
    minimum width=16.70cm,
    minimum height=2.35cm
  ] at (-0.15,2.80) {};
\end{scope}

\node[anchor=west, font=\bfseries\footnotesize] at (0.05,5.80)
  {Stage 1: Q-Former pretraining};

\node[frozen] (audio1) at (0.75,4.35)
  {Dialogue\\audio};

\node[frozen, minimum width=15mm] (tower1) at (2.30+0.25,4.35)
  {Frozen\\speech tower};

\node[frozen, minimum width=13mm] (emb1) at (3.85+0.4,4.35)
  {Speech\\emb.};

\draw[flow] (audio1) -- (tower1);
\draw[flow] (tower1) -- (emb1);

\node[
  qformer,
  minimum width=47mm,
  minimum height=25mm
] (qbox) at (6.55+1,4.35) {};

\node[font=\bfseries\tiny] at ($(qbox.north)+(0,0.25)$)
  {Clinical Q-Former (96 queries)};

\node[qS] (qs) at (5.25+0.55,4.95)
  {$Q_S$\\32};

\node[qO] (qo) at (5.25+0.55,4.35)
  {$Q_O$\\32};

\node[qAP] (qap) at (5.25+0.55,3.75)
  {$Q_{AP}$\\32};

\node[core] (qcore) at (6.55+0.95,4.35)
  {2$\times$ QF layer\\[-1mm]
   \tiny self-attn\\[-1mm]
   \tiny cross-attn\\[-1mm]
   \tiny FFN};

\node[qS] (zs) at (7.85+1.4,4.95)
  {$Z_S$};

\node[qO] (zo) at (7.85+1.4,4.35)
  {$Z_O$};

\node[qAP] (zap) at (7.85+1.4,3.75)
  {$Z_{AP}$};

\draw[sFlow]  (qs.east)  -- ($(qcore.west)+(0,0.38)$);
\draw[oFlow]  (qo.east)  -- (qcore.west);
\draw[apFlow] (qap.east) -- ($(qcore.west)+(0,-0.38)$);

\draw[sFlow]  ($(qcore.east)+(0,0.38)$) -- (zs.west);
\draw[oFlow]  (qcore.east) -- (zo.west);
\draw[apFlow] ($(qcore.east)+(0,-0.38)$) -- (zap.west);

\draw[kvflow] (emb1.east) -- node[above, note] {$K,V$} (qbox.west);

\node[
  lora,
  minimum width=14mm,
  minimum height=24mm
] (think1) at (9.70+1.65,4.35) {};

\node[font=\bfseries\tiny] at ($(think1.north)+(0,-0.27)$)
  {Thinker};

\node[note] at ($(think1.north)+(0,-0.52)$)
  {LoRA};

\coordinate (tinS)   at ($(think1.west)+(0,0.60)$);
\coordinate (tinO)   at ($(think1.west)+(0,0.00)$);
\coordinate (tinAP)  at ($(think1.west)+(0,-0.60)$);

\coordinate (toutS)  at ($(think1.east)+(0,0.60)$);
\coordinate (toutO)  at ($(think1.east)+(0,0.00)$);
\coordinate (toutAP) at ($(think1.east)+(0,-0.60)$);

\draw[sFlow]  (zs.east)  -- (tinS)  -- (toutS);
\draw[oFlow]  (zo.east)  -- (tinO)  -- (toutO);
\draw[apFlow] (zap.east) -- (tinAP) -- (toutAP);

\draw[
  black,
  line width=0.8pt
]
($(zo.east)!0.55!(tinO)$) ellipse [
  x radius=0.16cm,
  y radius=1.02cm
];

\coordinate (ellNorth) at ($(zo.east)!0.55!(tinO)+(0,0.92)$);
\draw[
  -Latex,
  black,
]
($(ellNorth)+(0.09,0)$) -- ($(think1.west)+(0,0.92)$);

\node[loss, minimum width=26mm] (lossGlobal) at (12.90+1.2,5.55)
  {global losses\\
   $\mathcal{L}_{SOAP}+\mathcal{L}_{concept}$};

\node[
  loss,
  draw=sColor,
  fill=sColor!10,
  text=sColor,
  minimum width=22mm
] (lossS) at (12.90+1,4.95)
  {$\mathcal{L}_S$\\Subjective};

\node[
  loss,
  draw=oColor,
  fill=oColor!10,
  text=oColor,
  minimum width=22mm
] (lossO) at (12.90+1,4.35)
  {$\mathcal{L}_O$\\Objective};

\node[
  loss,
  draw=apColor,
  fill=apColor!12,
  text=apColor,
  minimum width=22mm
] (lossAP) at (12.90+1,3.75)
  {$\mathcal{L}_{AP}$\\A\&P};

\draw[flow] ($(think1.east)+(0,0.92)$) -- ++(0.50,0) |- (lossGlobal.west);
\draw[sFlow]  (toutS)  -- (lossS.west);
\draw[oFlow]  (toutO)  -- (lossO.west);
\draw[apFlow] (toutAP) -- (lossAP.west);

\node[anchor=west, note] at (14.30+0.75,4.35)
  {section-specific\\auxiliary losses};

\node[anchor=west, font=\bfseries\footnotesize] at (0.05,2.55)
  {Stage 2: final SOAP tuning};

\node[frozen] (audio2) at (0.75,1.25)
  {Dialogue\\audio};

\node[frozen, minimum width=15mm] (tower2) at (2.30+0.25,1.25)
  {Frozen\\speech tower};

\node[frozen, minimum width=20mm] (proj2) at (4.40+0.5,1.70)
  {Frozen native\\audio proj.};

\node[
  qformer,
  fill=qformerFill!70,
  minimum width=20mm
] (qf2) at (4.40+0.5,0.80)
  {Frozen Stage-1\\Q-Former};

\node[loss, minimum width=28mm] (concat2) at (7.15+0.6,1.25)
  {Concatenate prefix\\
   Q-Former tokens first\\
   + native audio tokens};

\node[lora, minimum width=18mm] (think2) at (10.15+0.35,1.25)
  {Thinker\\new LoRA};

\node[loss, minimum width=17mm] (soap2) at (12.15+0.4,1.25)
  {SOAP\\note};

\node[loss, minimum width=17mm] (loss2) at (14.05+0.5,1.25)
  {$\mathcal{L}_{SOAP}$};

\draw[flow] (audio2) -- (tower2);

\draw[flow] (tower2.east) -- ++(0.35,0) |- (proj2.west);
\draw[flow] (tower2.east) -- ++(0.35,0) |- (qf2.west);

\draw[flow] (proj2.east) -- (concat2.west |- proj2.east);
\draw[oFlow] (qf2.east) -- (concat2.west |- qf2.east);

\draw[flow] (concat2) -- (think2);
\draw[flow] (think2) -- (soap2);
\draw[flow] (soap2) -- (loss2);

\end{tikzpicture}
\vspace{-0.1cm}
\caption{Overview of the two-stage section-aware Q-Former adaptation.
Stage 1 trains a clinical Q-Former with three SOAP-oriented query groups and
aligned section-specific auxiliary losses. Stage 2 freezes the Q-Former,
restores the native audio projection, and trains only a new thinker LoRA.}
\vspace{-0.4cm}
\label{fig:clean_qformer}
\end{figure*}


As shown in Figure \ref{fig:clean_qformer}, our system is built from the components of Qwen2.5-Omni-3B that are relevant to the BeTraC speech-to-SOAP summarization task. We remove the visual encoder and text-to-speech modules, and retain only the speech tower, the audio projection module, and the decoder-only language model, referred to as the thinker~\cite{xu2025qwen25omnitechnicalreport}.

A direct use of this architecture is not well suited to long medical dialogues. Such conversations produce a large number of audio tokens, many of which may encode information that is irrelevant. Feeding all these tokens directly to the thinker may increase the risk of hallucinated or repeated content. To address this limitation, we introduce a task-specific Q-Former adapter~\cite{dai2023instructblip}, trained to extract a compact set of clinically relevant representations from the speech embeddings. The system is trained in two stages.

\subsection{Q-Former pretraining}
In the first stage, we remove the native Qwen audio projector and replace it with a trainable Q-Former adapter. The speech tower remains frozen. The Q-Former receives the speech embeddings through cross-attention and produces a fixed set of learned query embeddings: 32 Subjective queries, 32 Objective queries, and 32 Assessment-and-Plan queries. This design encourages the adapter to organize the audio information according to the structure of the target SOAP note.

The Q-Former is trained jointly with a LoRA adapter~\cite{hu2022lora} added to the thinker. The training objective is to generate the SOAP note from the Q-Former output. In addition, to encourage the model to retain the factual clinical information from long conversations, we consider generating the list of medical concepts present in the reference note. These concepts are extracted from the ground-truth SOAP notes using the open medical matcher provided by the challenge organizers. We also add section-specific supervision: each query group is associated with an auxiliary loss that targets the corresponding SOAP section, namely Subjective, Objective, or Assessment and Plan. This stage therefore teaches the Q-Former to compress the audio dialogue into a small number of task-oriented clinical embeddings.

\subsection{Thinker final tuning}

In the second stage, we restore the native Qwen audio projector and integrate the pretrained Q-Former into the full system. The Q-Former weights learned in Stage 1 are kept frozen, and the LoRA adapter used in the first stage is discarded. A new LoRA adapter is then trained on the thinker, while the speech tower, the native audio projector, and the Q-Former remain frozen.

At this stage, the thinker receives two complementary sources of audio information. First, the frozen Q-Former produces a compact audio prefix containing 96 task-specialized clinical tokens. Second, the native Qwen audio pathway provides the projected speech embeddings generated by the original audio projector. The Q-Former tokens are concatenated before the native audio tokens and used as a clinical prefix for SOAP-note generation. This design allows the model to benefit from both the compressed, SOAP-oriented representation learned in Stage 1 and the richer acoustic information preserved by the native Qwen audio pathway. The only training objective in this stage is the generation of the complete SOAP note. The submitted model does not include diarization or explicit speaker-role masks and is trained fully end-to-end.

\subsection{Inference and post processing}

At inference time, we apply several decoding constraints to reduce hallucinations and repetitive outputs. We use beam search to improve generation quality and impose a maximum output length to avoid overly long responses. We also apply a repetition penalty~\cite{keskar2019ctrl} and block repeated n-grams in order to limit degenerate repetitions.

Finally, we apply a deterministic post-processing script to standardize the generated SOAP notes. This script removes special tokens and any text produced before the first Subjective section, normalizes section titles, subsection labels, and common medical terms using forms observed in the training SOAP notes, and removes duplicated content as well as irrelevant politeness formulas.

\section{Results} 
\subsection{Implementation details}
The submitted system is based on Qwen2.5-Omni-3B with a frozen speech encoder. The audio adapter is a Q-Former with 96 learned queries, 2 layers, 8 attention heads, and dropout 0.1. The thinker is adapted with LoRA (rank 16, $\alpha=32$). Training is performed on the official BeTraC training subset using one Nvidia H100, with gradient checkpointing, batch size 1, gradient accumulation 8, cosine learning rate decay, 5\% warmup, and 6 epochs. The LoRA learning rate is $5\times10^{-5}$, and the Q-Former learning rate in Stage 1 is $10^{-4}$. Model selection is based on the best validation loss.
At inference time, we use beam search with 16 beams, a maximum of 512 new tokens, a length penalty of 1.0, a repetition penalty of 1.2, and no-repeat 5-gram blocking.

\subsection{Challenge results}
\begin{table}[t]
\caption{Concept F1 on the different evaluation sets.}
\centering
\small
\begin{tabular}{l c c c c}
\hline
Team & Official\cite{labrak2026generating} & Mock\cite{Fareez2022} & Realistic & All \\
\hline
Top  & 0.543 & 0.505 & 0.534 & 0.531 \\
Our system & 0.501 & 0.455 & 0.425 & 0.486 \\
Lowest    & 0.362 & 0.347 & 0.305 & 0.357 \\
Baseline & 0.245 & 0.254 & 0.244 & 0.248 \\
\hline
\end{tabular}
\vspace{-0.2cm}
\label{tab:selected_f1}
\end{table}

\begin{figure}[ht]

    \centering
\includegraphics[width=0.9\linewidth]{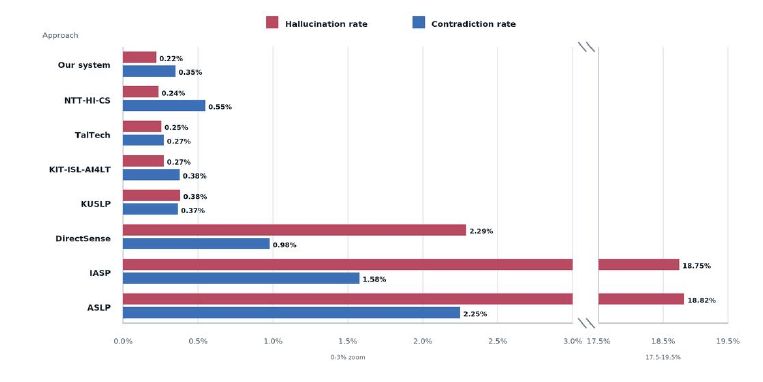}
    \caption{LLM-as-a-judge evaluation of the submitted systems.}
    \label{fig:llmasjudge}
\end{figure}

Our system ranked fourth in the lightweight track of BeTraC. The official ranking is based on SynthDoPaCo\cite{labrak2026generating} test set. On this set, our system reaches a Concept F1 of 0.5005, well above the challenge baseline and relatively close to the top system.
Table~\ref{tab:selected_f1} shows the same trend on additional evaluation sets released by the organizers. Our system remains clearly above the baseline on all subsets and reaches 0.486 when scores are aggregated. The largest drop appears on the Realistic subset, which may reflect lower robustness to human speech. It might also be explained by the fact that this subset contains only three dialogues.
The organizers reported an LLM-as-a-judge analysis. Our system scores below top systems on conciseness and structure, but remains closer on faithfulness and coverage. Moreover, it exhibits one of the best hallucination profiles in the challenge, as shown in Figure~\ref{fig:llmasjudge}.

\subsection{System analysis}

The LLM-as-a-judge analysis suggests that our system may be particularly robust to hallucinations. We therefore investigate this aspect further using deterministic heuristics, and compare the behavior of the different system variants under these indicators.
Table~\ref{tab:contrastive} compares four variants: the official Qwen-Omni challenge baseline, the same system with decoding constraints, the baseline fine-tuned and with decoding constraints, and our proposed system. Besides Concept F1, we report four heuristic indicators. \textit{Struct.} is the rate of notes with a valid SOAP structure. \textit{Claims} and \textit{Nums} are inspired by prior work on factual consistency such as FactCC\cite{kryscinski2020evaluating} and FActScore\cite{min2023factscore}. 
We compute lightweight unsupported-content diagnostics. Generated SOAP notes are decomposed into sentence-level claims and numeric expressions, which are then checked against the source dialogue and reference note. Claims or numbers with insufficient lexical evidence are counted as unsupported. For instance, an unsupported claim can be a generated diagnosis absent from both the dialogue and the reference note and an unsupported numerical value may be an invented blood pressure measurement. \textit{Rep.} captures degenerate repetitions. \textit{Hallu.} is the proportion of dialogues with at least one of the previous anomalies.

Two main effects emerge from Table~\ref{tab:contrastive}. First, decoding constraints mainly improve output form. They almost eliminate structural errors and repetitions, but do not improve factual reliability. In particular, unsupported numbers remain a major issue. Second, compared with the fine-tuned Qwen-Omni baseline, our proposed system improves both factual reliability and task performance. As shown in Table~\ref{tab:contrastive}, it yields fewer unsupported claims, fewer unsupported numbers, and a lower overall hallucination rate, while also reaching the best Concept F1. This indicates that the benefit of our method is not limited to format or scoring, but also improves the grounding of generated notes.
\begin{table}[t]
\caption{Progressive comparison from the baseline to our full system using Concept F1 and hallucination-related metrics.}
\centering
\small
\resizebox{\linewidth}{!}{
\begin{tabular}{l c c c c c c}
\hline
Method & C-F1 \textuparrow & Struct \textuparrow & Claims \textdownarrow & Nums \textdownarrow & Rep \textdownarrow & Hallu \textdownarrow\\
\hline
Baseline & 0.26 & 47\% & 9.7\% & 15.0\% & 13.0\% & 26.5\% \\
+Dec. constr. & 0.28 & 98.5\% & 9.4\% & 60.0\% & 0.0\% & 60.5\% \\
+Fine-tuned & 0.47 & 100\% & 1.3\% & 15.5\% & 0.0\% & 15.5\% \\
+Proposed & 0.51 & 99.8\% & 0.4\% & 7.5\% & 0.0\% & 7.5\% \\
\hline
\end{tabular}
}
\vspace{-0.2cm}
\label{tab:contrastive}
\end{table}
We also examined more adverse audio conditions in the validation set. We identified two dialogues with more than two minutes of consecutive non-verbal content and nine dialogues with degraded speech. On this latter subset, the fine-tuned baseline degrades (0.47 to 0.31 C-F1), whereas our proposed system remains much more stable (0.51 to 0.48).

\begin{figure}[ht]
    \centering
    \includegraphics[width=0.9\linewidth]{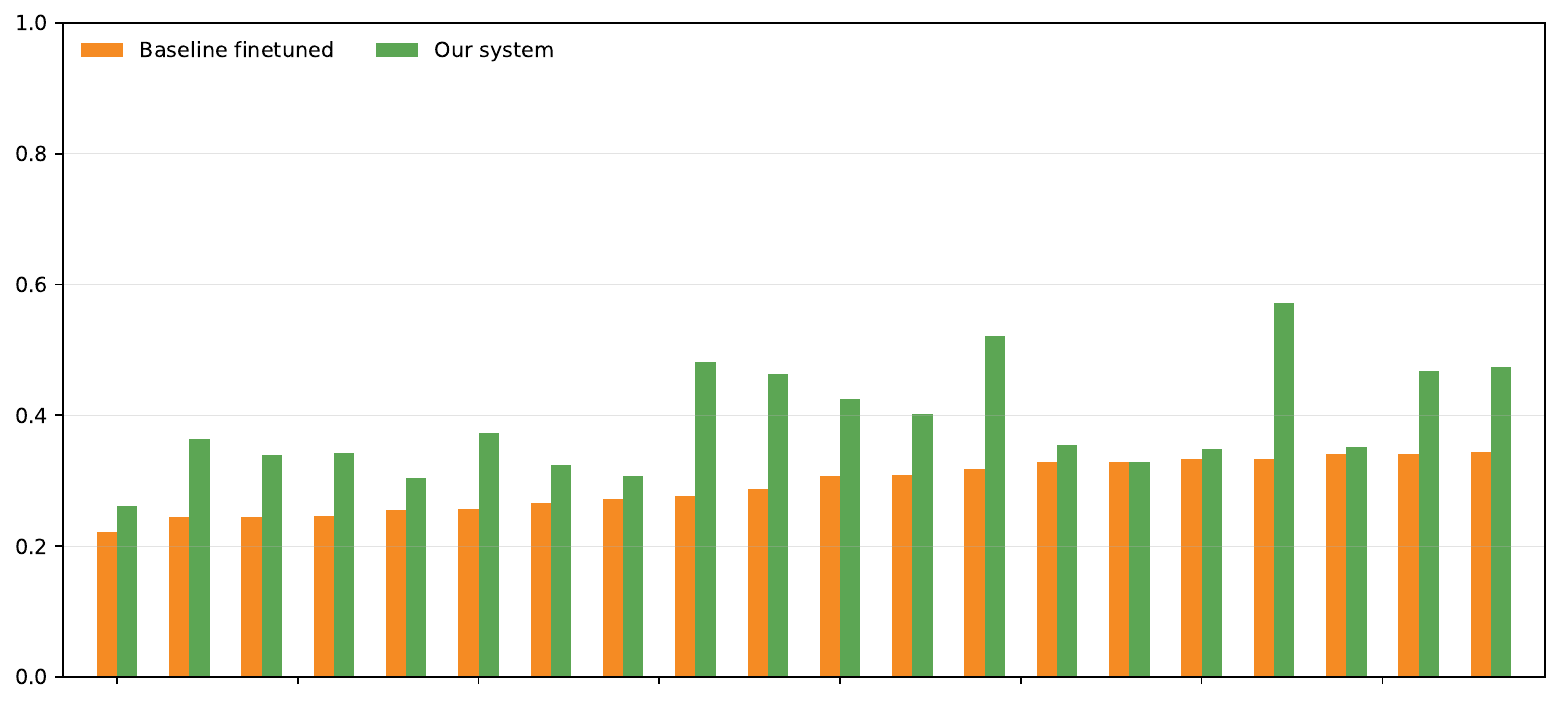}
    \caption{Concept F1 on the worst 5\% dialogues of the fine-tuned baseline.}
    \label{fig:dialogwise_f1}
\end{figure}

Finally, Figure~\ref{fig:dialogwise_f1} further confirms the contribution of our method on the weakest cases of the fine-tuned Qwen-Omni baseline. The improvement is consistent across all selected dialogues. More generally, our system wins against this baseline for 77.5\% of validation dialogues. Overall, our proposed system seems to improve both factual reliability and Concept F1, with the largest gains on the most challenging dialogues.

\section{Conclusions}
We presented a system for medical note generation from doctor-patient conversations. It substantially improves over the challenge baseline and ranked fourth in the BeTraC lightweight track. Our method relies on SOAP section-specific queries to extract information for each part of the note. This design likely encourages better grounding in the source conversation. This is consistent with both the organizers' LLM-as-a-judge analysis and our heuristic study, suggesting improved robustness to hallucinations.

\clearpage
\bibliographystyle{unsrt}
\bibliography{mybib}

\end{document}